# The MAJORANA $^{76}$Ge neutrino less double-beta decay project: A brief update

**Frank T. Avignone III**
Department of Physics and Astronomy, University of South Carolina, Columbia, South Carolina 29208, USA
(On behalf of the MAJORANA Collaboration)

E-mail: avignone@sc.edu

**Abstract.** At present, MAJORANA is a research and development (R&D) project to investigate the feasibility and cost of constructing and operating a one ton $^{76}Ge$ $0\nu\beta\beta$–decay experiment with ~1000 kg of Ge detectors fabricated from germanium enriched to 86% in $^{76}Ge$. The study will include three separate cryostats with various types of detectors: un-segmented, un-segmented point-contact, minimally segmented, and highly segmented. One cryostat will contain at least 30 kg of enriched (preferably point-contact) detectors. The performance of the cryostats and detectors as well as background levels will be investigated. The goal of the demonstrator project is to reach a $3\sigma$ discovery sensitivity of $\sim 10^{26}$ y.

## 1. Introduction

The subject of double –beta decay has been well covered in several recent review articles [1-5]. It is well known from available experimental evidence from neutrino oscillation experiments [6], that at least one of the neutrino mass eigenvalues is approximately 0.05 eV. In the case of the inverted-mass hierarchy $(m_3 \approx m_2 >> m_1)$, the solar oscillations involve $(m_3 \Leftrightarrow m_2)$, while atmospheric oscillations involve $(m_{3,2} \Leftrightarrow m_1)$. An approximate expression of the theoretical rate can be written as:

$$(T_{1/2}^{0\nu})^{-1} = G^{0\nu}(E_0,Z)\left|\frac{\langle m_{\beta\beta}\rangle}{m_e}\right|^2 \left|M_f^{0\nu} - (g_A/g_V)^2 M_{GT}^{0\nu}\right|^2. \qquad (1)$$

If the average oscillation parameters are used, one can write the following expression for the effective Majorana mass of the electron neutrino, in case of the inverted hierarchy:

$$\langle m_{\beta\beta}\rangle = \left|(0.70^{+0.02}_{-0.04})m_3 + (0.30^{+0.04}_{-0.02})m_2 e^{i\phi_2} + (\leq 0.05)m_1 e^{i(\phi_3+\delta)}\right|. \qquad (2)$$

The atmospheric oscillation data [7] imply that $m_3 - m_1 \approx 0.05 eV$. With various nuclear structure calculations of the quantity: $\left|M_f^{0\nu} - (g_A/g_V)M_{GT}^{0\nu}\right|$, and of the phase-space factor, $G^{0\nu}(E_0,Z)$, one can predict the half-life corresponding to $\langle m_{\beta\beta}\rangle \approx 0.05 eV$, which is the case of the inverted hierarchy when the lightest eigenstate mass, $m_1$, is near zero.

Three very recent nuclear structure models were used to compute the half-lives corresponding to $\langle m_{\beta\beta}\rangle \approx 0.05 eV$. They yield the following values for the $0\nu\beta\beta$–decay of $^{76}Ge$: $3.2 \times 10^{27} y$ (Shell Model) [8], $(0.6 - 1.2) \times 10^{27} y$ (RQRPA) [9], and $1.5 \times 10^{27} y$ (QRPA) [10]. The longest of these should be the target sensitivity of a one-ton $^{76}Ge$ $0\nu\beta\beta$–decay experiment.

## 2. The MAJORANA Demonstrator R&D project

The goal of the project is to design and build a "Demonstrator Module" containing about 60 kg of germanium, of which 30 kg will be enriched to 86% or more in $^{76}Ge$. The two most sensitive $0\nu\beta\beta$–decay results reported thus far are for $T_{1/2}^{0\nu}(^{76}Ge)$, and are from the Heidelberg-Moscow experiment: $1.9 \times 10^{25} y$ [11], and the IGEX experiment: $1.6 \times 10^{25} y$ [12]. The one-ton experiment must increase the sensitivity by about a factor of 100. There is an unconfirmed claim of observation at $(1.30 - 3.55) \times 10^{25} y$ $(3\sigma)$ [13]. Both the MAJORANA Demonstrator and the GERDA Phase-1 experiments are designed to be able to confirm or rule out that result.

Germaium-76 $0\nu\beta\beta$–decay searches have several advantages: isotopic enrichment is well established; Ge detectors use a well-developed technology that can include pulse-shape discrimination and segmentation; they have the best energy resolution of any of the proposed techniques (~0.16% at 2039-keV) i.e., at the $\beta\beta$–decay end–point energy of $^{76}Ge$. They can also take advantage of the lessons learned from the Heidelberg-Moscow [11], and IGEX [12] experiments.

MAJORANA, in cooperation with the GERDA collaboration [14], is involved in R&D aimed at developing a one-ton scale $0\nu\beta\beta$–decay experiment. The specific MAJORANA goals are:
- To build a prototype module to test the claim of discovery of $0\nu\beta\beta$–decay [13],
- To demonstrate a background low enough to justify a 1-ton experiment,
- To prepare for a down-select between MAJORANA and GERDA techniques, and
- To pursue longer term R&D to minimize cost and to optimize the schedule for constructing the optimum one-ton scale $^{76}Ge$ $0\nu\beta\beta$–decay experiment.

The reference-design parameters of the prototype Demonstrator Module that will demonstrate the efficacy of the design and will meet the science goals are as follows:
- Approximately 60-kg of ~1-kg crystals in a close-packed configuration.
- At least 30 kg of the Ge will be enriched to 86% or more in $^{76}Ge$,
- A likely mix of detector types (n-type, p-type, segmented, point contact etc.),
- A module design scalable to a one-ton scale experiment, with the cryostat constructed of ultra-low background electroformed copper as in IGEX, and
- Located deep underground (~4000 mwe) and enclosed in a low background passive shield surrounded by an active veto shield.

An experiment with 30-kg of Ge enriched to 86% in $^{76}Ge$, producing data for a live time of 3 years, or approximately 77-kg.y of $^{76}Ge$ exposure, with a background of 1 count in the region of interest per ton-year, should have a sensitivity of approximately $T_{1/2}^{0\nu}(^{76}Ge) \geq 10^{26} y$. This will probe the entire $3\sigma$ range of the discovery claim $(1.30 - 3.55) \times 10^{25} y$ of reference [13]. See Figure 1.

The main questions that should be answered by the Demonstrator Module are as follows:
- Is the copper low enough in radioactive backgrounds; will it be necessary to electroform the copper parts underground?
- Can the small parts and cables be made pure enough to meet background requirements?
- Can large cryostats be operated with the detectors cooled by radiation?
- What will be the optimum detector configuration, point-contact detectors, modestly segmented, or highly segmented?
- How well does segmentation reject background?

The time evolution of the sensitivity of the 30-kg of enriched $^{76}Ge$, operating for a live time of three years, is shown in Figure 1. The darkened region represents the $3\sigma$ claimed discovery range $(1.30-3.55)\times 10^{25} y$, corresponding to the most recent analysis of Klapdor-Kleingrothaus and Krivosheina [13].

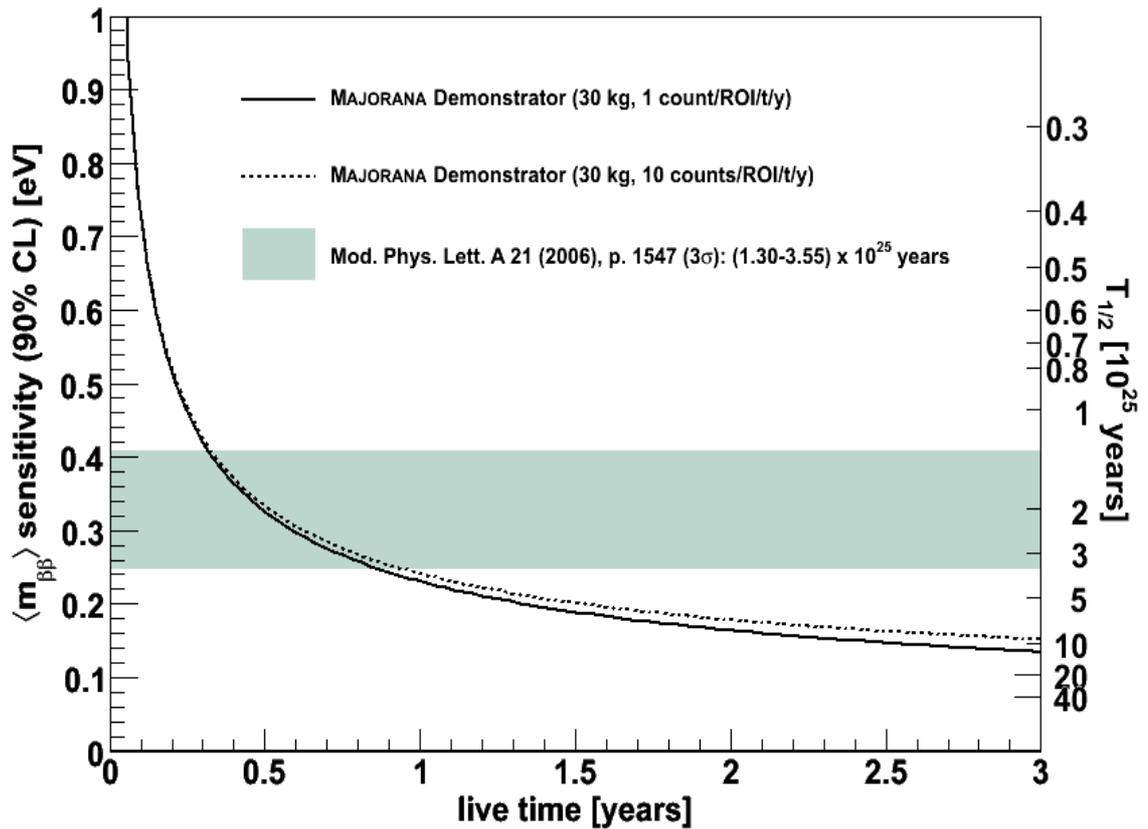

**Figure 1.** The sensitivity achievable with 30 kg of enriched Ge in the MAJORANA Demonstrator Module in the configuration discussed in the text.